\documentclass[11pt,article]{IEEEtran}

\usepackage{amsmath,amsthm,amssymb}
\usepackage{thmtools}
\usepackage{amsfonts}
\usepackage{graphicx}
\usepackage{float}
\usepackage[square,numbers]{natbib}
\usepackage{dirtytalk}
\usepackage{multirow}
\usepackage{caption} 

\PassOptionsToPackage{hyphens}{url}\usepackage{hyperref}

\author{Michael Filletti}
\date{May 2019}
\title{What influence has Brexit had on financial markets, in particular GBP/EUR exchange rates?}

\begin{document}

\maketitle

\section{Introduction} \label{sec:intro}
%\textit{What is the conclusion of your task study and what evidence (if any) do you have to support your conclusion. The rest of the task section outlined below will discuss how you arrived to this conclusion.}
%\\
%\newline
On 23rd June 2016, 51.9\% of British voters voted to leave the European Union (EU), triggering a process and events that will lead to the United Kingdom leaving the EU, an event that has commonly become known as 'Brexit'. On 29 March 2017, then Prime Minister Theresa May formally triggered Article 50, which began a two year countdown that should have led to Brexit occurring on 29 March 2019. Since then, the process has been anything but smooth, and the Brexit date has in fact been extended to October 2019 as at time of writing. In this assignment, we investigate the effects of this entire process on the currency markets, specifically the GBP/EUR exchange rate. Financial markets are known to be quite sensitive to unexpected news and events, and the relationship between financial exchange rates and news has been investigated in many papers \citep{bauwens2005,jin2013,nassirtoussi2015}. Despite being a relatively recent event, the impact of Brexit on financial markets has also been looked into by various authors \citep{tabeshian2018,bousselmi2018}. In her work, \citet{tabeshian2018} investigates whether Brexit was a shock for the currency and stock markets, using abnormal returns and the ARCH/GARCH model to study the effects. \citet{bousselmi2018} look into the long term market performance of British and European firms, investigating how Brexit influences this. Both studies look into the abnormal returns for the various firms and look into the short and long term effects. Abnormal Returns are used by multiple other authors when investigating the impact of Brexit \citep{burdekin2018,oehler2017,schiereck2016,tielmann2017}.
\\
\newline
%\textit{\textbf{Our findings and conclusions}}
%\\
%\newline
To answer the question posed, we follow a similar path to that of the previously mention works of \citet{tabeshian2018}, and incorporate some of the methodology used by \citet{mahajan2008}. Rather than investigating a specific date, like \citet{tabeshian2018} does, we investigate the entire timeline and effects to date, to see how the situation has evolved. The first step is to obtain the data, which involves data from news agencies and financial exchange rate data. Once the data is cleaned and in a useful format, we find the number of articles related to Brexit published as well as the number of times the term Brexit is mentioned in the articles each day. Our methodology is based on the assumption that the more news articles are released and the more a term is mentioned, the more likely that a major event occurred. \citet{jin2013} attempt a very similar task, which is to identify and analyse major events in depth, based on a similar assumption. We then look into the correlation between the number of times a term is mentioned and the exchange rate. This helps us understand when certain events are impacting the exchange rate. As an alternative approach, the days at which a spike in the number of articles published is seen are considered major events, and the exchange rate is investigated on those days. Once the major events are identified, the abnormal returns of the exchange rate are found for those days. A numeraire is used to be sure that the drop is coming from the GBP currency, and not the EUR. Furthermore, we investigate if various cryptocurrencies may have been influenced by Brexit events. Finally, Latent Dirichlet Allocation (LDA) is run to further investigate the event date.
\section{Data Collection} \label{sec:coll}
%\textit{Describe the data collection process in detail, e.g. the data source(s) used, sampling frequency, etc..}
For this task, we require two types of data, news article data and currency exchange rate data. Currency exchange rate data is more readily available than news article data, with various R Software packages providing this. One of these packages is \textbf{Quandl}, which allows the user to obtain whatever data one may require via their API. To use the Quandl function, the user is simply required to input their API key (obtained when signing up to Quandl), as well as the code to which the database belongs. In the scope of this task, the following Quandl data sources are used:
\begin{itemize}
    \item ECB/EURGBP
    \item ECB/EURUSD
    \item CHRIS/ICE\_MP1
    \item BITFINEX/XRPUSD
    \item BITFINEX/BTCUSD
    \item BITFINEX/LTCUSD
\end{itemize}
Despite our main interest being in the GBP/EUR conversion rate, we also collect data on the USD rate to act as a numeraire, which means that it acts as a control to show that any drops in the GBP/EUR rate is coming from one currency and not the other. All the above datasets provide data at a daily level, which is what was required for the task carried out. Note that the \textit{ECB} (European Central Bank), \textit{CHRIS/ICE\_MP1} and \textit{BITFINEX} datasets all have different formats and return various variables. The \textit{ECB} provides the data for each trading day, and contains data dating back from 1995. The official site of the \textit{ECB}\footnote{https://www.ecb.europa.eu/} claims that rates are usually updated at 16:00 CET. The \textbf{FinancialScript.R} script is used to obtain the data, and stores data for all conversion rates in separate CSV files.
\\
\newline
The collection of news article data is far less straightforward. Various large news agencies were considered, namely the Associated Press, the Guardian and the Telegraph, however, scraping attempts were unsuccessful. The scraping of the Reuters site for headlines was successful through the \textbf{rvest} package and the \textbf{NewsScraperv2.R} script. Brexit related headlines were found in the following sections:
\begin{itemize}
    \item Brexit
    \item British Politics
    \item British Economy
\end{itemize}
The major issue with the data from these sources is that it does not cover the entirety of our timeline. News headlines from the Brexit section only goes as far back as November 2017, while those of the British Politics section begin at 29 June 2016, meaning that we would not have headlines related to the buildup to the referendum of 24 June 2016, or those concerning the actual referendum. The British Economy section on the other hand, was far too small to return any significant data.
\\
\newline
For this reason, an alternative more consistent source of data was still required. Using the \textbf{BeautifulSoup} package of Python within the \textbf{Python ToM Scraper.ipynb} file, the Times of Malta site was successfully scraped, both for headlines and articles. This data was obtained through the search function of the Times of Malta site, and by using packages the urllib to access the various pages and articles. Because of the structure of the site, the urllib package was required to input certain parameters such as the search keyword (which was set to be Brexit) and the time period  (set to be between 01/01/2016 and 20/04/2019. Once these were set, the search ID was found using the BeautifulSoup parser, as well as the number of pages the search contains. The search URLs could be set up, using the parameters, a cookie (which matches a Search ID) and the article pages. Despite the Times of Malta being a relatively minute news agency, it is the best data that was obtained, so we make (and test) the assumption that the pattern and behaviour of the publication of news articles resembles that of the largest agencies (such as Reuters and PA).
\section{Data Cleansing} \label{sec:clean}
%\textit{Describe the data cleansing techniques used, e.g. duplicate removal, normalisation of data, which data you decided to use (filtering) and why, etc.}
%\\
%\newline
The exchange rate data requires just two variables, the \textit{Date} and the \textit{Value}. The \textit{Value} was taken to be the mean of the exchange rate for the day. Certain more technical values such as the highest or lowest values of the rate on a specific day were not considered. The \textit{Settle} column is used from the \textit{CHRIS/ICE\_MP1} dataset, and the \textit{Mid} column is used from the \textit{BITFINEX} dataset, as these represent the average of the day. The main hurdle when dealing with FOREX data is how to deal with weekends. Once again, the \textbf{FinancialScript.R} script cleans the data, after having collected it (as mentioned in section \ref{sec:coll}. Naturally, during weekends there is no rate to report, however news articles are still ongoing. For this type of situation there are two approaches, or scenarios, that one can consider. The first is to run the analysis excluding weekends entirely, considering five day weeks \citep{andersen2003,laakkonen2004,yermack2015}, both for the news articles and for exchange rates. An alternative approach is to take the rates on the Saturday and Sunday to be the same as those on Friday, as the rates see little to no change on those days, as claimed by \citet{muller1990}. Due to the fact that a notable amount of articles are still published on the weekends, and that these may influence our correlation with the exchange rate, the decision is made to exclude weekends entirely from our analysis.
\\
\newline
The Times of Malta article data contains the following variables:
\begin{itemize}
    \item \textit{Title}
    %\item \textit{Author}
    \item \textit{Article}
    \item \textit{URL}
    \item \textit{Date}
    \item \textit{Year}
\end{itemize}
The \textit{URL} is of the form \textit{\url{https://www.timesofmalta.com/articles/type/YYYYMMDD/restofURL}}, meaning that it could be used to obtain the date, and convert it to the following form: \textit{DD/MM/YYYY}. The \textbf{trimTOMurl} function in the \textbf{ToMPreProcessing.R} script is used to extract the date from the URL and construct the date. Every article has text (such as information on terms and conditions and copyright text) that must be removed. The \textbf{trimTOMarticle} function, also found in the aforementioned script, is used to exclude such text, and the article is trimmed to only contain the relevant text. Once the article has been trimmed, text pre-processing is carried out, which involves the following steps:
\begin{enumerate}
    \item Converting text to lower case
    \item Removing non-ASCII text and punctuation
    \item Tokenizing
    \item Removing stopwords
    \item Removing numbers
\end{enumerate}
These steps are carried out through the \textbf{Clean\_String}, \textbf{Remove\_Stopwords} and \textbf{Remove\_Numbers} functions, within the \textbf{ToMPreProcessing.R} script. The above text preprocessing steps are also carried out for the Reuters dataset, through the \textbf{ReutersPreProcessing.R}. No text trimming is required for this dataset, however some irrelevant headlines had to be removed. This processing is required for us to have the ability to further investigate what topics are popular on certain dates, by carrying out a Latent Dirichlet Allocation (LDA).
\section{Data Storage} \label{sec:store}
%\textit{Describe the data storage and model, e.g. how the data was saved prior to analysis, etc.}
For the scope of this task, as mentioned in section \ref{sec:clean}, the average exchange rate for each day is used, while data is required from \textit{1 January 2016} to \textit{19 April 2019}. Below are the variables of our final financial dataset:
\begin{itemize}
    \item \textit{Date}
    \item \textit{Value} - Average exchange rate for that specified date
\end{itemize}
We have multiple financial datasets, related to traditional currencies and the conversion rates between them (\textit{EUR}, \textit{USD}, \textit{GBP}), and also containing the rate of the \textit{USD} to multiple cryptocurrencies (\textit{XRP}, \textit{BTC}, \textit{LTC}). The raw \textit{GBP/EUR} and \textit{EUR/USD} datasets simply contain the date and average rate for that day. The other datasets are more detailed, containing the peak and low of the rate for the day, the average, and other variables. The raw datasets are each stored in their own CSV file.
\\
\newline
News from Times of Malta was stored in a CSV file named \textit{news5.csv}, containing the \textit{Title}, \textit{Article}, \textit{Author}, \textit{Year} and \textit{URL} of every Brexit related publication between \textit{1 January 2016} to \textit{19 April 2019}. Preprocessing was carried out using \textbf{ToMPreProcessing.R} as explained in section \ref{sec:clean}, and stored in \textit{FinalDataToM.csv}. The Reuters data was stored in three separate datasets, named \textit{DataReutersBP.csv} (Reuters British Politics),\textit{DataReutersBE.csv} (Reuters British Economy) and \textit{DataReutersSE.csv} (Reuters Brexit). Once the datasets are preprocessed, and non-Brexit related stories are filtered out by the \textbf{ReutersPreProcessing.R} script, they are each sent into another group of CSV files, \textit{FinalDataReutersBP.csv}, \textit{FinalDataReutersBE.csv} and \textit{FinalDataReutersSE.csv}. Using the \textbf{DataMerge.R} file, the datasets are transformed to fit one dataset, named \textit{BrexitNewsData.csv}. This dataset contains the following variables:
\begin{itemize}
    \item \textit{Date} - Date article published
    \item \textit{Article} - Article text assigned, for Reuters data this is the headline along with the description
    \item \textit{Tokens} - Cleaned text, obtained from the article column
    \item \textit{Source} - Source of news article (either Reuters or Times of Malta)
\end{itemize}

\section{Data Analysis} \label{sec:analysis}
%\textit{Evidence for the selected strategy (this includes statistical analysis, visualisations, descriptive statistics, modelling including assumptions taken, hypothesis testing, etc.)}
%\\
%\newline
Once the data is stored and available, the analysis can be addressed. The first question that is to be to answered is in relation to the correlation between major Brexit events and currency fluctuations. There are multiple methods to approach this question, however, in our approach we focus on using the article data as an integral part of our solution. The major assumption that we make is to assume that when a major event occurs, the number of articles, and by extension the number of times the term \textit{Brexit} is mentioned, go up. This is based on the approach made by \citet{mahajan2008}, who tackle a similar problem to us, analysing financial news to identify major events. Using LDA, they find periodic highs of these topics, and correlate them with the fluctuations in stock prices. At moments of high correlation, this is indicates a strong likelihood of a major event occurring and influencing the exchange rate. Our methodology is quite similar. By using the number of times the term \textit{Brexit} is used per day, or alternatively the amount of articles published, we find moments at which \textit{Brexit} was being highly reported, and consequently, had major events happening. The decision on whether to use the number of articles published, or the number of times Brexit was mentioned is not a straightforward decision.
\\
\newline
The next step is to investigate the correlation between the number of times \textit{Brexit} is mentioned against the GBP/EUR conversion rate. The Pearson Correlation is used to measure the strength of relationship between one metric variable and another. The correlation is a value within $[-1,1]$, with -1 representing the strongest possible negative linear relationship, while 1 represents the strongest possible positive linear relationship. On the other hand, a value of 0 indicates that there is no relationship. Equation \ref{PCC} below shows how the correlation is calculated, where $\mathbf{X}$ and $\mathbf{Y}$ represent two variables containing values $X_{1},\ldots,X_{n}$ and $Y_{1},\ldots,Y_{n}$ respectively.
\begin{equation} \label{PCC}
    r_{\mathbf{XY}}=\frac{\sum^{n}_{i=1}(X_{i}-\bar{X})(Y_{i}-\bar{Y})}{\sqrt{\sum^{n}_{i=1}(X_{i}-\bar{X})^{2}}\sqrt{\sum^{n}_{i=1}(Y_{i}-\bar{Y})^{2}}}
\end{equation}

Once again, there are multiple methods and approaches of obtaining this correlation. The three methods used are:
\begin{itemize}
    \item Number of \textit{Brexit} Mentions/Articles vs Exchange Rate
    \item Number of \textit{Brexit} Mentions/Articles vs \% Change of Exchange Rate
    \item Number of \textit{Brexit} Mentions/Articles \% Change of Exchange Rate
\end{itemize}
These comparisons have been used by \citet{mahajan2008} in their study to investigate the correlation of topics and stock prices. The next step is to decide how to take the correlation. As we're taking a huge amount of time (over 3 years of data), once again, we consider multiple approaches:
\begin{itemize}
    \item Overall Correlation - Finding the correlation of the two series over the entire time
    \item Cumulative Correlation - Investigating how the correlation develops as time goes by
    \item Window Correlation - Investigating the correlation between the two series over a specific window of 10 days
\end{itemize}
Obtaining the dates at which there is a major event is one of the more difficult tasks in this assignment. For this reason, we once again consider multiple scenarios, to investigate and evaluate them. The first approach is based on what was described in Window Correlation above, and is based on the approach of \citet{mahajan2008}. Using window correlation, we are able to identify windows (dates) at which the number of times the term \textit{Brexit} is mentioned, correlates significantly with the GBP/EUR exchange rate. The days at which the correlation is above 0.602 (50\% significance with 9 degrees of freedom), are selected to be major events. Naturally, for days that are within each others windows (for example the dates 01/01/1990, 02/01/1990 and 03/01/1990), one of the dates is selected and investigated further for significance. The dates are also filtered, so as to only return dates when the number of articles published is above average. This is to avoid for any scenarios of no articles being released, and the exchange rate remaining stable, as this would clearly not be an event of interest to us. The second approach is far more simple and straightforward. In this approach we select the days which had a top 5\% number of articles released. In this scenario, we apply our major assumption, that more articles are published when a major event occurs, far more bluntly. In both cases, a list of dates is obtained. These dates are further investigated in the context of the exchange rate, by using the abnormal returns method. 
\\
\newline
The Abnormal Returns is a method used to carry out an event study. It has been used to study the impact of the Brexit referendum \citep{burdekin2018,tabeshian2018}. However, in this assignment, we aim to extend this analysis and use Abnormal Returns to investigate if there are any other dates that experienced a significant drop. \citet{mackinlay1997} writes a paper discussing this methodology and several approaches that can be taken. To test an for abnormal returns, the first step is to define the event of interest. In our scenario, this will have been defined through our two methodologies, that were the window correlation and the days that were in the top 5\% of articles released. Once these are obtained, we define the estimation and event window. The estimation window is the period of time within which we find the expected rate of the exchange rate, while the event window is the window within which the event occurs. The length of the estimation window varies, with some authors using 100 days \citep{cox1994}, while others use 250 days \citep{mackinlay1997}. The event window is naturally a much shorter time range, with \citet{mackinlay1997} using a $[-1,1]$ range, while, \citet{tabeshian2018} uses up to $[-5,5]$ days as the event window. There are many ways to estimate the performance during the estimation window. In this study, we use the more straightforward and simple Mean Adjusted Return model, described by \citet{mackinlay1997}. The Abnormal Returns is calculated using equation \ref{AbRet}, shown below.
\begin{equation} \label{AbRet}
    \textrm{AR}_{t}=\textrm{R}_{t}-\mathbb{E}[\textrm{R}_{t}|X]
\end{equation}
where $\mathbb{E}[R_{t}|X]=\frac{1}{T}\sum^{T}_{i=1} \textrm{R}_{i}$, such that $T$ is the length of the event window. %In addition to $\textrm{AR}_{t}$, the cumulative abnormal returns $\textrm{CAR}$ may also be used:
%\begin{equation} \label{CAR}
%    \textrm{CAR}=\sum_{t=T_{1}}^{T_{2}}\textrm{AR}_{t}
%\end{equation}
Once the abnormal returns has been obtained, the next step is to obtain some test statistics. By using the t-test, we test the following hypotheses:
\begin{flalign*} \label{hyp:a} 
    H_{0}: \textrm{AR}_{t} =  0 &&\\
    H_{1}: \textrm{AR}_{t} \neq 0 &&
\end{flalign*}
The $t$-statistic for the $A\mathrm{AR}$ is calculated using equation \ref{tstatar} shown below:
\begin{equation} \label{tstatar}
    t_{AR}=\frac{\textrm{AR}_t}{s_{AR}}
\end{equation}
where $s_{AR}^{2}=\frac{1}{(T_{2}-T_{1})-2}\sum_{t=T_{1}}^{T_2}(\textrm{AR}_{t}^{2})$. %Similarly, the $t$-statistic for the $\mathrm{CAR}$ is shown in equation \ref{tstatcar}:
%\begin{equation} \label{tstatcar}
%    t_{CAR}=\frac{\textrm{CAR}_t}{s_{CAR}}
%\end{equation}
%where $s_{CAR}^{2}=(T_{2}-T_{1}) \cdot s_{CAR}^{2}$.

\begin{figure}[h!]
\centering
\includegraphics[scale=0.2]{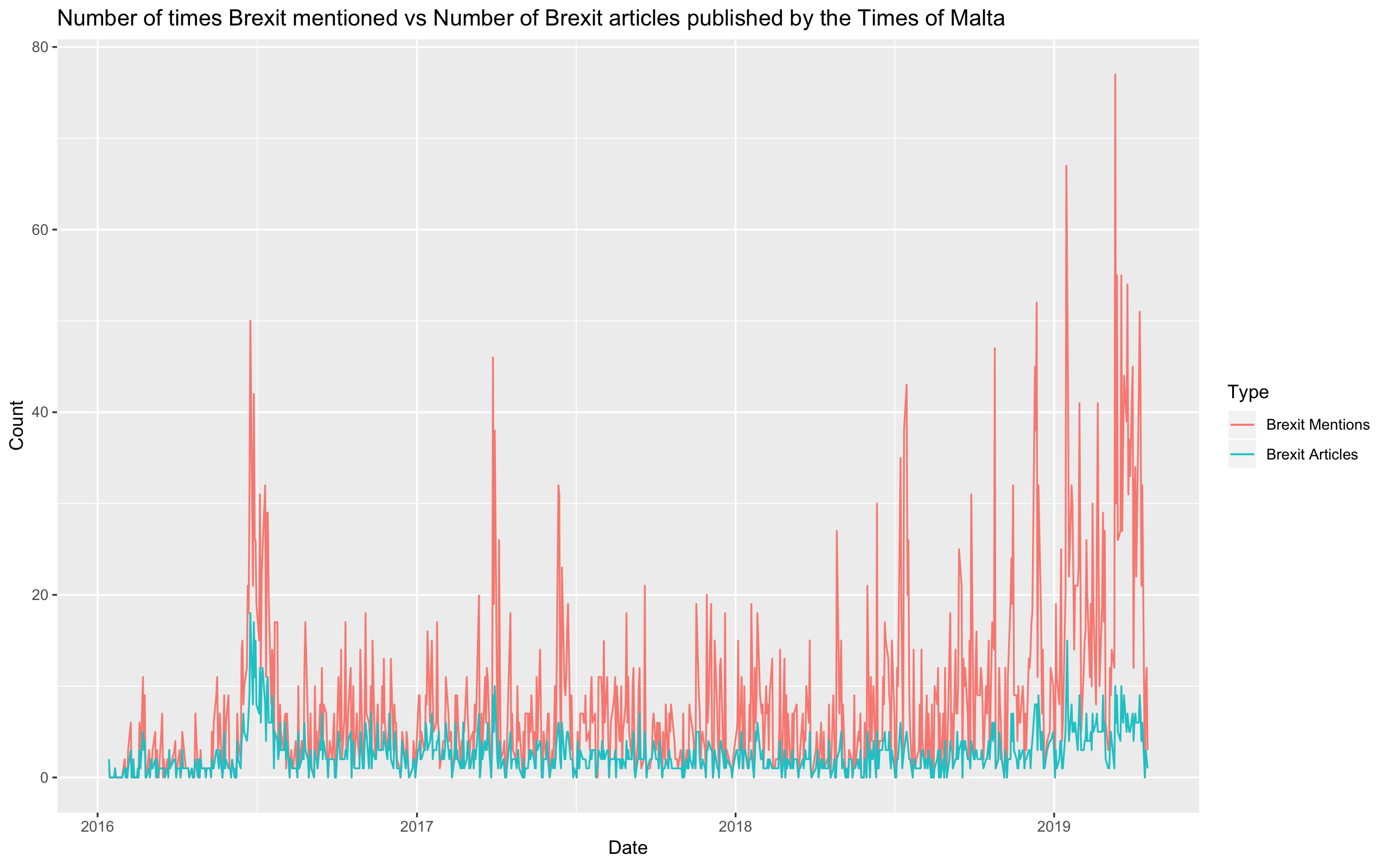}
\caption{Graphs illustrating the number of articles published against the number of times Brexit was mentioned in the articles}
\label{fig:MentvsArt}
\end{figure}

Once the dates at which the abnormal returns have shown a significant change have been identified, we run an LDA. The LDA model consists of words, documents, which are a sequence of words, and an entire corpus, which is a collection of documents. Each document is seen as a multinomial distribution of multiple topics, while each topic is a multinomial distribution of words. By estimating the parameters of these two distributions, we are able to identify the topics belonging to our corpus. LDA is used to obtain topics for those specific dates, and build somewhat of a timeline in relation to Brexit. Certain topics are selected to illustrate how the situation has changed, by showing when these topics formed a significant part of Brexit articles.
\\
\newline
We consider 3,020 articles from the Times of Malta, between 1 January 2016 and 19 April 2019. On average, 2.42 articles are released per day. The highest number of times that articles related to Brexit have been published in a day is 22 times, which happened the day after the 2016 referendum, on 25th June 2016. On average, the term \textit{Brexit} is mentioned 2.8 times per article, with a variance of 3.9. This is a very high variance, and not ideal, as it indicates a strong amount of noise. However, upon looking at the correlation between the two series, which is 0.779, an extremely strong and high correlation, this indicates that despite there being a high level of noise, the number of times \textit{Brexit} is mentioned correlates strongly to that of the number of articles. Figure \ref{fig:MentvsArt} compares the number of articles published and the number of times Brexit was mentioned. There is a seemingly close correlation, confirmed by the correlation figure mentioned previously.
Table \ref{Corr1} illustrates the various approaches taken to investigate the overall correlation between the news data and GBP/EUR exchange rate. When it comes to the raw exchange rate, there is a noticeable negative correlation between the number of times \textit{Brexit} is mentioned and the rate. There is also a strong correlation between the absolute percentage change in the exchange rate and the number of articles released. There is no noticeable correlation in other approaches. For the purpose of this assignment, we carry out further analyses by using the \textit{Brexit} mentions along with the raw exchange rate. We investigate the cumulative correlation to investigate how this correlation evolves over time. Figure \ref{fig:CummCorr} below shows that there is a particular period of time at which the correlation is noticeably negative, and overall tends to lean in that fashion.
\begin{table}[H]
\centering
\begin{tabular}{ |p{4cm}||c|c| }
    \hline
         \textbf{Type of correlation}& \textbf{Brexit Mentions} & \textbf{Articles} \\ \hline
    \hline
        \textbf{Exchange Rate} & -0.142 & -0.080 \\ \hline
        \textbf{\% Change in Exchange Rate} & -0.029 & -0.059 \\ \hline
        \textbf{Absolute \% Change in Exchange Rate} & 0.078 & 0.160  \\ \hline
\end{tabular}\par
\caption{Correlation of news data against GBP/EUR exchange rate over entire time period}
\label{Corr1}
\end{table}

\begin{figure}[h!]
\centering
\includegraphics[scale=0.2]{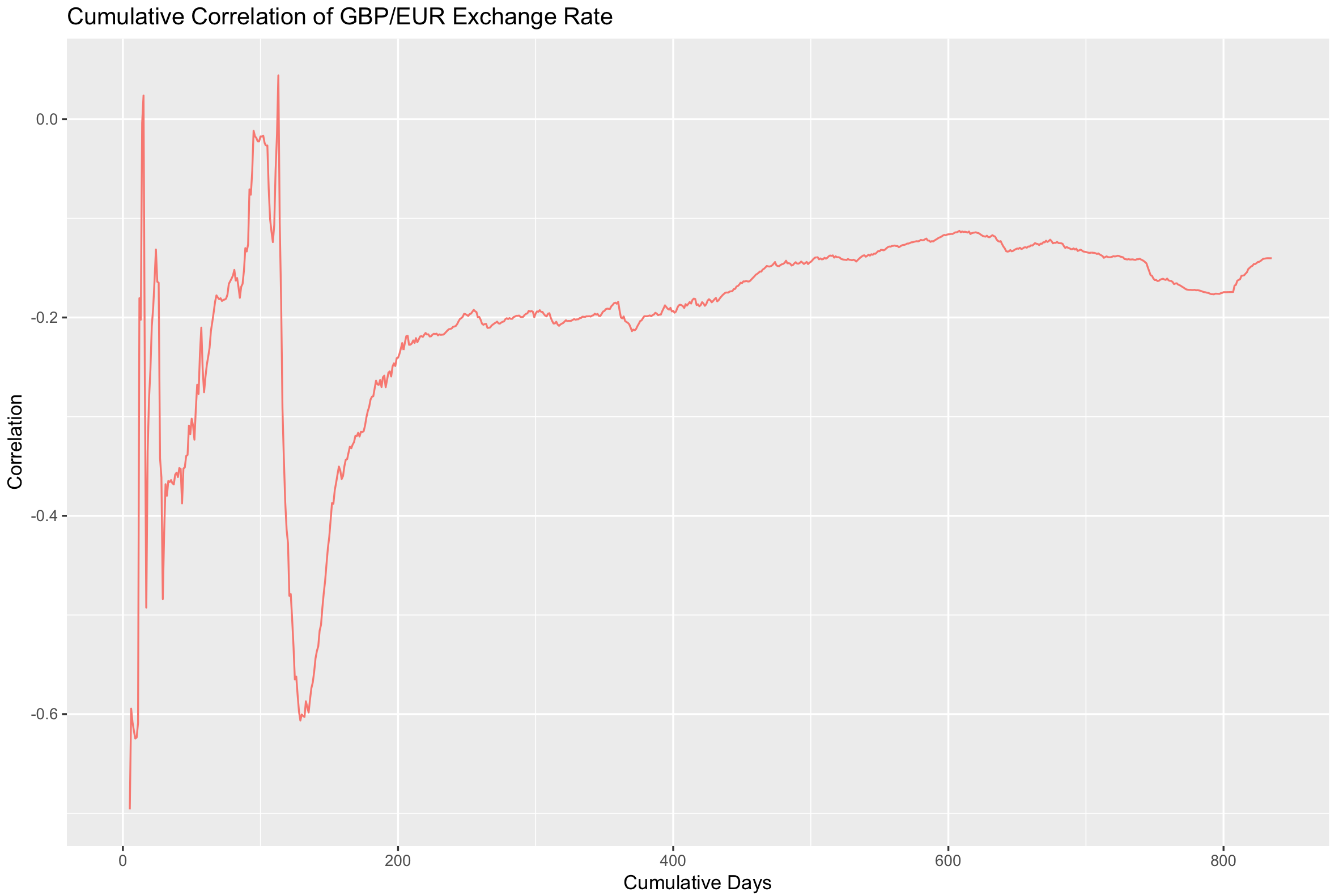}
\caption{Cumulative Correlation of Exchange Rate across entire time period}
\label{fig:CummCorr}
\end{figure}

\begin{figure}[h!]
\centering
\includegraphics[scale=0.2]{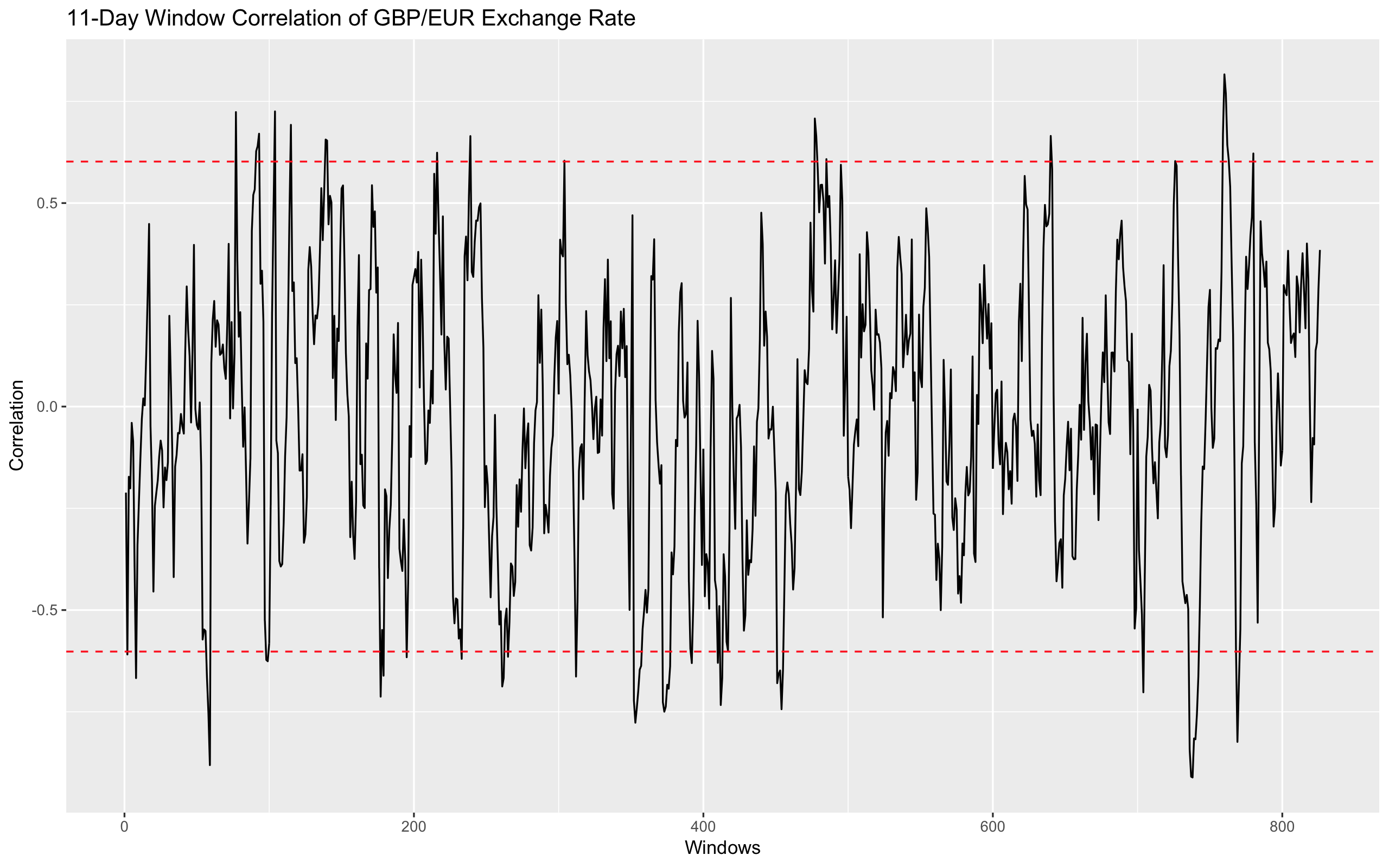}
\caption{Eleven day Window Correlation of Exchange Rate across entire time period}
\label{fig:WindCorr}
\end{figure}

Window Correlation involves finding the correlation of the two series in spaces of eleven days. Eleven days are chosen as the event window is of the same length. This should help us in identifying periods within which the exchange rate and news both fluctuate. As has been previously mentioned, this is an approach based on that of \citep{mahajan2008}. Figure \ref{fig:WindCorr} illustrates the correlation across the various windows and days considered. The red dashed line represents the critical value at which the correlation is considered to be significant.
The \textbf{WindowCorrelation} function that returns this graph also returns the dates which exist within the window that have a significant correlation. Since eleven days are considered here, this means we use 9 degrees of freedom. With 95\% significance, the critical value for significant correlation is 0.602, as indicated on the graph. As we are aiming to identify major events, we look for dates when the number of articles published is above the average. This is used to indicate to us that there is a strong likelihood that a major Brexit-related event occurred and it influenced the exchange rate. One of the major event dates obtained from this approach is 24th June 2016, the day after the Brexit referendum. This is to be expected, as it is a day that significantly impacted the whole currency market \citep{tabeshian2018}.
\begin{table}[H]
\centering
\begin{tabular}{ |c||c|c|c|c| }
    \hline
         \textbf{Date} & \textbf{Time} & \textbf{GBP/EUR} & \textbf{Abnormal Return} & $\mathbf{t}_{\textrm{AR}}$  \\ \hline
    \hline
        \textbf{17/06/2016} & -5 & 1.270 & 0.004 & 0.630 \\ \hline
        \textbf{20/06/2016} & -4 & 1.292 & 0.019 & \textbf{2.914} \\ \hline
        \textbf{21/06/2016} & -3 & 1.303 & 0.009 & 1.367 \\ \hline
        \textbf{22/06/2016} & -2 & 1.302 & 0.000 & 0.034 \\ \hline
        \textbf{23/06/2016} & -1 & 1.306 & 0.003 & 0.513 \\ \hline
        \textbf{24/06/2016} & 0 & 1.238 & -0.050 & \textbf{-7.888} \\ \hline
        \textbf{27/06/2016} & 1 & 1.199 & -0.031 & \textbf{-4.829} \\ \hline
        \textbf{28/06/2016} & 2 & 1.209 & 0.009 & 1.389 \\ \hline
        \textbf{29/06/2016} & 3 & 1.211 & 0.003 & 0.431 \\ \hline
        \textbf{30/06/2016} & 4 & 1.210 & 0.000 & -0.077 \\ \hline
        \textbf{01/07/2016} & 5 & 1.193 & -0.013 & \textbf{-2.077} \\ \hline
\end{tabular}\par
\caption{Abnormal Returns obtained for 24th June 2016}
\label{tab:AR1}
\end{table}
Table \ref{tab:AR1} illustrates the abnormal returns results for 24th June 2016, taking 150 days to be the estimation window, and 11 days as the event window. Dates that see a significant abnormal return (at 90\% significance), are shown in bold. It is clear that there is an enormous drop in the GBP/EUR rate on 24th June, with $\textrm{AR}=-0.050$ on the day, and $\textrm{AR}=-0.031$ the following day. In addition, we see another significant drop on 1st July 2016. Just four days before the referendum, there was a significant increase in the rate, however this is more likely to be part of the volatility due to the uncertainty that came with the referendum.
\\
\newline
Upon running the above exercise for event windows within which there was a strong correlation between news and financial data, and days that had an above average number of articles, we obtain a list of days that saw a 95\% significant level of change in the GBP/EUR exchange rate. Table \ref{tab:AR2} shows the abnormal returns for the dates identified. It is interesting to note that the GBP/EUR and GBP/USD both tend to fluctuate in a similar way, whilst the same cannot be said for the GBP/EUR and EUR/USD. This drives us toward the notion that the currency that is experiencing the changes is the GBP currency, rather than the EUR currency.
\begin{table}[H]
\centering
\begin{tabular}{ |c||c|c|c| }
    \hline
         \textbf{Date} & \textbf{GBP/EUR} & \textbf{GBP/USD} & \textbf{EUR/USD}  \\ \hline
    \hline
        \textbf{24/05/2016} & \textbf{0.013} & 0.01 & -0.004 \\ \hline
        \textbf{01/06/2016} & \textbf{-0.015} & -0.004 & 0.002 \\ \hline
        \textbf{06/06/2016} & \textbf{-0.018} & -0.004 & \textbf{0.017} \\ \hline
        \textbf{20/06/2016} & \textbf{0.018} & \textbf{0.024} & 0.007 \\ \hline
        \textbf{24/06/2016} & \textbf{-0.051} & \textbf{-0.078} & \textbf{-0.029} \\ \hline
        \textbf{27/06/2016} & \textbf{-0.032} & \textbf{-0.035} & -0.007 \\ \hline
        \textbf{07/10/2016} & \textbf{-0.022} & -0.013 & -0.004 \\ \hline
        \textbf{01/12/2016} & \textbf{0.014} & 0.007 & 0 \\ \hline
        \textbf{09/12/2016} & \textbf{0.013} & -0.001 & \textbf{-0.018} \\ \hline
        \textbf{25/01/2017} & \textbf{0.012} & 0.011 & 0 \\ \hline
        \textbf{18/07/2017} & \textbf{-0.011} & -0.001 & 0.007 \\ \hline
        \textbf{21/07/2017} & \textbf{-0.01} & 0.002 & \textbf{0.013} \\ \hline
        \textbf{13/10/2017} & \textbf{0.015} & 0.001 & -0.005 \\ \hline
        \textbf{29/11/2017} & \textbf{0.014} & 0.004 & -0.006 \\ \hline
        \textbf{29/11/2018} & \textbf{-0.01} & -0.004 & \textbf{0.009} \\ \hline
        \textbf{10/12/2018} & \textbf{-0.013} & \textbf{-0.015} & 0.005 \\ \hline
        \textbf{31/12/2018} & \textbf{0.009} & 0.004 & 0 \\ \hline
        \textbf{02/01/2019} & \textbf{-0.008} & -0.01 & -0.004 \\ \hline
        \textbf{14/01/2019} & \textbf{0.008} & 0.001 & -0.006 \\ \hline
        \textbf{23/01/2019} & \textbf{0.009} & 0.008 & 0.001 \\ \hline
\end{tabular}\par
\caption{Significance of Abnormal Returns obtained for various exchange rates and major events identified by Window Correlation approach}
\label{tab:AR2}
\end{table}
To evaluate the strength of these results, we compare the identified days in figure \ref{tab:AR2} to those of the Brexit timeline outlined by \citet{walker2017}. The four dates identified before the referendum, highlighted as having significant changes and a strong correlation between news and financial data, are likely to be down to the volatility of the rate at the time, due to uncertainty. In fact, on these four dates, the rate both increases and decreases. The identified dates of 24th and 27th June 2016 are clearly an effect of the referendum that took place, and were the most evident examples of such a correlation between a Brexit event and a change in the exchange rate. On 2nd October 2016, then Prime Minister Theresa May confirmed that Article 50 will be triggered before the end of March 2017. The event date identified by our approach is the 7th October 2016, when we see a significant drop in the GBP/EUR rate. During the first few days of December 2016, the Government appealed against the High Court that Article 50 may be triggered without the vote of Parliament. The exchange rate saw an increase in the GBP/EUR rate, however we should note that the EUR/USD rate decreased, meaning that this increase may also have come due to the EUR currency losing value.
\\
\newline
The days between 17th January 2017 and 26th January 2017 contain some significant events, one of which was the Supreme Court rejecting the appeal made by the Government. Throughout these events, the GBP/EUR exchange rate saw some statistically significant increases, however was still unable to recover to the original value. The next significant change identified comes during mid-July. These dates are a few days after the Government introduced the EU Withdrawal Bill. During these days, the currency saw statistically significant drops, once again. The next dates identified that also some significant events came at the end of 2018. During November 2018, the Withdrawal Agreement made notable steps forward, being agreed and published, and was endorsed by EU27 leaders. Following this, the rate showed another statistically significant drop on 29th November 2018. However, It should be noted that on the same day, the EUR/USD rate saw a significant increase, meaning that the change could be due to the EUR increasing in value. During December 2018, the Prime Minister announces a delay on the Meaningful Vote, that was supposed to occur the following day. This decision seemingly had an instantaneous effect, with the GBP losing statistically significant value the very next day. The next major events identified came in January 2019, showing a small but statistically significant increase in the rate. Interestingly, during this month, on 15th January, the Prime Minister lost the Meaningful Vote, resulting in the leader of the Opposition placing a motion of no confidence in the Government.
\\
\newline
As is clear, certain events of Brexit clearly have impacted the GBP/EUR currency rate. The value of the GBP currency has seen an evident drop between 24th June 2016 to this day, and has not recovered. Certain events pushing towards Brexit occurring caused a statistically significant drop in the rate, whilst the vote of no confidence that occurred in January 2019, seems to have had a positive effect on the rate. This methodology captured many of the major events, however also missed out on some big events that occurred, particularly in the first half of 2017 and 2018. It should also be noted that by using the USD as a numeraire, we were successfully able to identify changes in the rate that may not have been linked to Brexit, but due to the EUR changing in value.
\begin{table}[H]
\centering
\begin{tabular}{ |c||c|c|c| }
    \hline
         \textbf{Date} & \textbf{GBP/EUR} & \textbf{GBP/USD} & \textbf{EUR/USD}  \\ \hline
    \hline
        \textbf{20/06/2016} & \textbf{0.018} & \textbf{0.024} & 0.007 \\ \hline
        \textbf{24/06/2016} & \textbf{-0.051} & \textbf{-0.078} & \textbf{-0.029} \\ \hline
        \textbf{27/06/2016} & \textbf{-0.032} & \textbf{-0.035} & -0.007 \\ \hline
        \textbf{01/07/2016} & \textbf{-0.014} & 0.004 & 0.003 \\ \hline
        \textbf{10/12/2018} & \textbf{-0.013} & \textbf{-0.015} & 0.005 \\ \hline
        \textbf{14/01/2019} & \textbf{0.008} & 0.001 & -0.006 \\ \hline
        \textbf{23/01/2019} & \textbf{0.009} & 0.008 & 0.001 \\ \hline
        \textbf{26/02/2019} & \textbf{0.009} & \textbf{0.013} & 0.001 \\ \hline
        \textbf{20/03/2019} & \textbf{-0.009} & -0.002 & 0 \\ \hline
        \textbf{22/03/2019} & \textbf{0.009} & 0.01 & \textbf{-0.007} \\ \hline
\end{tabular}\par
\caption{Significance of Abnormal Returns obtained for various exchange rates and major events identified by Top 5\% Articles approach}
\label{tab:AR3}
\end{table}
\begin{table}[H]
\centering
\begin{tabular}{ |c||c|c|c| }
    \hline
         \textbf{Date} & \textbf{LTC/USD} & \textbf{XRP/USD} & \textbf{BTC/USD}  \\ \hline
    \hline
        \textbf{24/05/2016} & -0.002 & NA & 0.008 \\ \hline
        \textbf{01/06/2016} & 0.017 & NA & 0.014 \\ \hline
        \textbf{06/06/2016} & 0.023 & NA & 0.012 \\ \hline
        \textbf{24/06/2016} & \textbf{0.08} & NA & 0.042 \\ \hline
        \textbf{27/06/2016} & 0.007 & NA & 0.021 \\ \hline
        \textbf{07/10/2016} & 0.007 & NA & 0.007 \\ \hline
        \textbf{01/12/2016} & 0.007 & NA & 0.01 \\ \hline
        \textbf{09/12/2016} & 0.006 & NA & -0.001 \\ \hline
        \textbf{25/01/2017} & -0.018 & NA & -0.025 \\ \hline
        \textbf{18/07/2017} & -0.003 & 0.061 & NA \\ \hline
        \textbf{21/07/2017} & 0.008 & -0.026 & NA \\ \hline
        \textbf{13/10/2017} & 0.032 & 0.016 & 0.052 \\ \hline
        \textbf{29/11/2017} & 0.03 & \textbf{0.131} & 0.008 \\ \hline
        \textbf{29/11/2018} & -0.029 & -0.039 & -0.003 \\ \hline
        \textbf{10/12/2018} & -0.045 & -0.031 & -0.031 \\ \hline
        \textbf{31/12/2018} & -0.052 & -0.057 & -0.033 \\ \hline
        \textbf{02/01/2019} & 0.06 & 0.033 & 0.03 \\ \hline
        \textbf{14/01/2019} & 0.075 & 0.055 & 0.049 \\ \hline
        \textbf{23/01/2019} & 0.006 & -0.013 & -0.003 \\ \hline
\end{tabular}\par
\caption{Significance of Abnormal Returns obtained for various exchange rates and major events identified by Window Correlation approach}
\label{tab:AR4}
\end{table}
Upon using the dates that are within the top 5\% of articles published daily, far fewer dates that show a significant change are 12th March 2019, the Prime Minister lost the second Meaningful Vote, and on 20th March, we see another significant drop on one of our outlined dates.
\\
\newline
The cryptocurrency results are investigated with the Window Correlation dates identified, and we investigate if any major changes have occurred throughout. Interestingly, on the date that Brexit happened, LTC increased with statistical significance, however, this is most likely to be a coincidence. Despite having such a correlation there is unlikely to be any form of causation there. This change can also be seen for BTC, and given our results in figure \ref{tab:AR2}, this increase is more likely to be due to a drop in the USD currency. In fact, changes in the rates occurred very rarely together with major Brexit events, meaning that these cryptocurrencies seem to be fairly unaffected by the events. Table \ref{tab:AR4} illustrates the obtained results.
\\
\newline
Finally, we attempt to construct some sort of timeline based on the dates obtained from our approaches. By using LDA we are able to extract topics from our documents. When running LDA on specific dates we obtain a mix of words representing topics. By considering the top 20 words forming the topics for each date, we select the following relevant words:
\begin{itemize}
    \item \textit{brexit}
    \item \textit{deal}
    \item \textit{delay}
    \item \textit{referendum}
    \item \textit{theresa}
    \item \textit{vote}
\end{itemize}
When these words are seen to form part of the top 20 words of a collection of documents published within the specified event window, we plot them in the our graph in figure \ref{fig:LDA}. When we speak of top 20, we are referring to the $\beta$ parameter, which is the probability of the word for that specific topic. For the purpose of this task, we make use of Reuters news data in conjunction with Times of Malta data, so as to get more words and documents within the event window. Figure \ref{fig:LDA} illustrates to us how the situation changes over time. As one can see, during June 2016 the \textit{referendum} is one of the more popular words, along with the \textit{vote} word, which seems quite popular throughout the whole timeline. From mid-2017 onwards, the \textit{deal} topic also becomes one of the more popular topics. Towards the end of our timeline, the word \textit{delay} has become prominent. This illustrates in a simple way how the topics of conversation have changed throughout the entire process.
\begin{figure}[h!]
\centering
\includegraphics[scale=0.2]{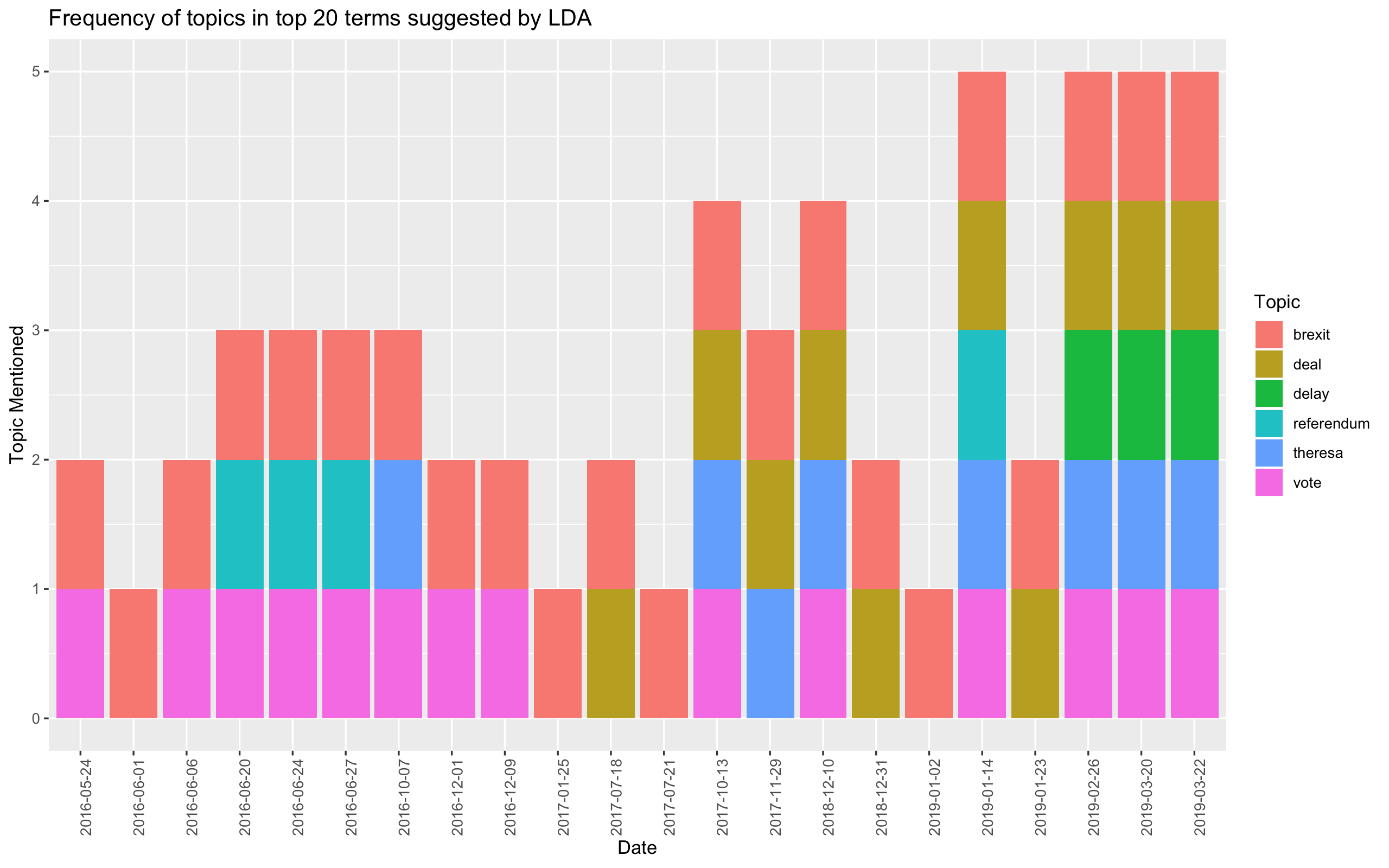}
\caption{Plot showing some common words representing topics for various identified Brexit dates}
\label{fig:LDA}
\end{figure}
\section{Conclusion} \label{sec:conc}
%\textit{Critical review of your solution. Also, a discussion of shortcomings and possible improvements is required. The Future Work section should not be banal, overly generic or marginal.}
Brexit has undeniably had a significant effect on the exchange rate, this is common knowledge, and something that has been confirmed discussed in multiple papers \citep{bousselmi2018,tabeshian2018}. In our approach, we look into multiple events, and see that in some cases the correlation still exists, and the effects are real. The timeline created provides a further confirmation that certainly, some part of our approach brings results that are clear and make sense. By simply using the number of times a term was mentioned and the number of articles released, from a relatively small newspaper, we were able to build a timeline and identify major events, and their correlation with the fluctuations of the exchange rate. However, despite this, there are clearly many shortcomings within our approach. One of the most notable and significant shortcomings is in relation to the news data. Despite the results showing that the Times of Malta can provide a fairly acceptable simulation of the general news landscape, it is still an extremely small newspaper, and much stronger results would have been yielded had sites such as the Guardian, the Associated Press and the Times been successfully scraped. Obtaining more news sources would also have helped the performance of the Top 5\% approach used to obtain the dates. With a larger dataset, the approach would have most likely returned far stronger results. Lastly, and most importantly, it would have strengthened and given more credence to our general assumption that when major events occur, more articles are published. An alternative to our assumption that news articles spike when a major event occurs is to try and attempt to apply Timeline Summarization from the headlines. Timeline Summarization could be used to make use of headlines and cluster them, finding the Influence, Spread and Informing values, as discussed by \citet{tran2015}. This is a fairly complex methodology that makes use of Random Walks and Logistic Regression, however returned positive results, so could be further investigated and tested within this scenario.
\\
\newline
The Abnormal Returns is a simple, yet effective method of identifying significant changes within a time series. Another common methodology to analyse a time series is through ARCH/GARCH modelling. This method was used in conjunction with Brexit by \citet{tabeshian2018}, who used the technique to discover that prior to the referendum there was a significant amount of volatility. This may have been interesting to see what the volatility was like in the build up to certain events, such as the build up to the multiple Meaningful Votes that were carried out during early 2019. An interesting exercise to further build upon what was done in this assignment may also be to investigate how certain terms correlate with the change in the exchange rate. Perhaps when a certain term is often used, the currency tends to fluctuate in a certain way. There are a vast number of improvements that can be made to this approach. However, the positive results indicate a strong foundation for more detailed and accurate approaches to be undertaken, and further investigate the effect that Brexit is having on the markets.

\newpage

\bibliographystyle{abbrvnat}
\bibliography{ics5115.bib}

\end{document}